# SEISMIC EFFECT OF GROUP WINNING BLASTING – CASE STUDY FROM A POLISH COPPER MINE

**Piotr Mertuszka**[1]**, Krzysztof Fuławka**[1]**, Lech Stolecki**[1]**, Marcin Szumny**[1],

[1]KGHM CUPRUM Ltd. Research and Development Centre, **Poland**

**ABSTRACT:**

*The group winning blasting is one of the most often applied method of high-energy tremors provoking in underground mines of the Lower Silesian Copper Basin. The simultaneous detonation of dozens mining faces are performed in order to release of elastic energy which is cumulated in surrounding rock mass. The accuracy of currently used non-electric initiation systems is insufficient to provide a controlled interference of seismic wave generated by blasting works. As a result, simultaneous detonation of higher number of mining faces does not always correlate with improvement of the effectiveness of rockburst prevention. In this paper, the records of seismic waves generated by blasting works in one of the mining panels from the Polish copper mine were analysed. The evaluation was performed by correlation of selected blasting parameters such as the number of fired mining faces and the total amount of detonated explosives with parameters of induced seismic wave.*

## 1. INTRODUCTION

The copper ore deposit located in underground mines of the Lower Silesian Copper Basin lies at depths reaching even 1,500 m. Under such conditions, mining activity generates high pressures within the rock mass which in turn are the cause of damage to workings. As the mined out area increases, the risk of seismic activity does so also, often leading to rockburst and roof falls (Fuławka et al., 2018; Saharan & Mitri, 2011). In order to reduce the risk of rockburst occurrence, a number of preventive actions are applied in the form of active and passive prevention methods. The former are mainly of an organisational nature, therefore their effectiveness should be considered in the long term. On the other hand, active methods, in great simplification are based on the destress of rock mass by detonation of explosives. This methods can be treated as "ad hoc" and at the same time the most effective. The basic method of seismic events provoking in underground mines of the Legnica-Głogów Copper Basin is group winning blasting which is based on simultaneous firing of group of mining faces. In result it may lead to release of the seismic energy accumulated in the rock mass (Butra & Kudełko, 2011; Wojtecki et al., 2017).

Furthermore, as Kabiesz and Lurka (2014) pointed out, it is expected, that simultaneous firing of multiple mining faces may amplify purposely the energy of induced seismic vibrations. To obtain constructive vibration interference from blasting, it is necessary to maintain proper synchronization of time delays during the firing of subsequent faces (Kabiesz et al., 2015). Unfortunately, the current initiation systems are not sufficiently precise to achieve a controlled local amplification of the seismic waves generated by the detonation of explosives. This means that the simultaneous firing of explosive charges in a larger number of mining faces does not always improve the effectiveness of rockburst prevention.

In this study, the records of seismic vibrations generated by firing of group of mining faces in the conditions of a selected mining panel of the Polish copper mine were analysed. All records were examined in terms of the possibility of amplification of paraseismic vibrations and the effectiveness of seismic event provocation. The assessment of the seismic effect of group blasting was based on the correlation of selected parameters of blasting works (number of fired faces and total amount of explosives applied) with the determined parameters of seismic vibrations (duration, maximum amplitude and dominant frequency of vibration).

## 2. SEISMIC ACTIVITY

With the progress of mining within considered mining panel, an increase of seismic activity, mainly in the form of high-energy tremors was observed. In recent years, starting from the third quarter of 2014, seismic activity has increased almost linearly, both in number and energy of tremors, what is presented in Figure 1.

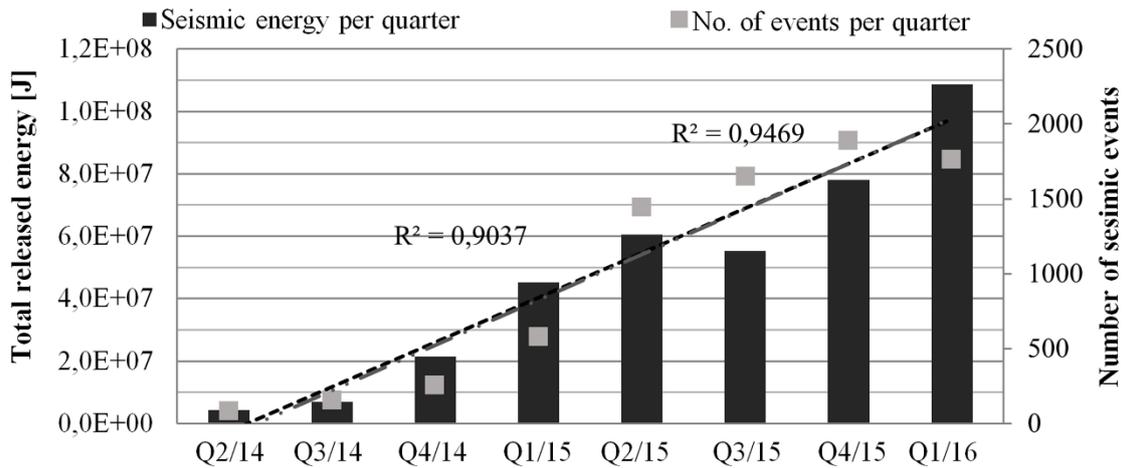

Figure 1. Event count and energy release trends between 2014 and 2016

Therefore, in order to minimise the risk of an accident, active methods of rockburst prevention in the form of group blasting were applied. Some seismicity, which is not connected with the underground production blasting may be considered as spontaneous, while provoked seismicity means, that it occurs in the time of firing or within so-called waiting time, i.e. time immediately following the production blasting.

The effectiveness of the high-energy tremors provocation was maintained at a relatively low level both in quantitative and energetic terms during the period considered. From the time of occurrence point of view, most of the observed seismic events are of random nature, even though that group blasting operations were frequently applied. Moreover, an increase in the number of provoked events in the time period immediately following blasting (when the crew is outside hazard zones) was not observed. One of the reasons may be the insufficient energy impulse generated by detonation of explosives.

3. ASSESSMENT OF THE SEISMIC EFFECT OF GROUP WINNING BLASTING

For the purpose of analysis, the multi-face group blasting carried out between May 2016 and December 2016 were studied. There were 140 multi-face blasting conducted within considered mining panel during this period (Figure 2).

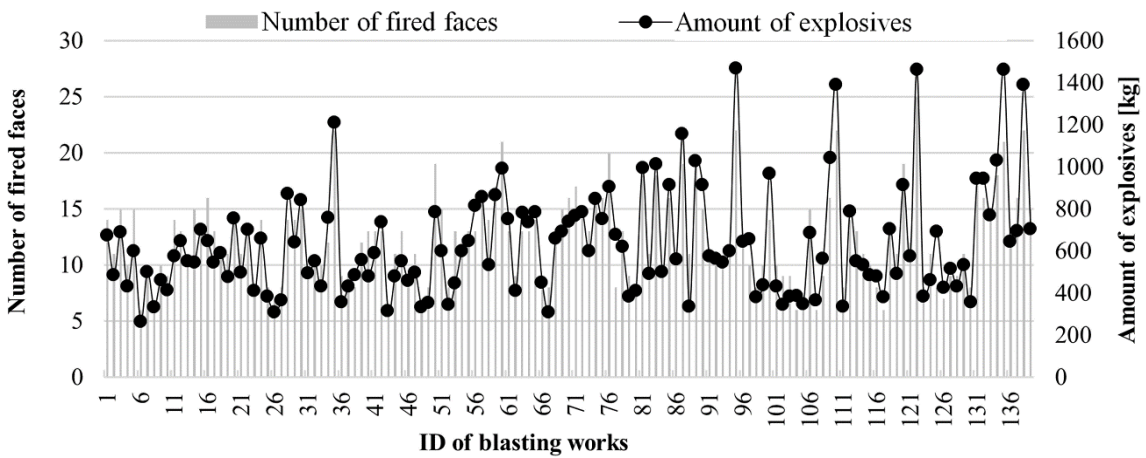

Figure 2. Summary of multi-face blasting operation within considered period

From abovementioned blasting works, 17 events were selected and analysed, differing in the number of simultaneous fired mining faces and amount of explosives. The list of group blasting works selected for the analysis of the seismic effect are presented in Table 1.

Table 1. List of multi-face blasting operations

| Blasting ID | 1 | 2 | 3 | 4 | 5 | 6 | 7 | 8 | 9 | 10 | 11 | 12 | 13 | 14 | 15 | 16 | 17 |
|---|---|---|---|---|---|---|---|---|---|---|---|---|---|---|---|---|---|
| Number of fired mining faces | 9 | 9 | 8 | 14 | 10 | 10 | 11 | 14 | 14 | 15 | 18 | 18 | 22 | 22 | 21 | 25 | 22 |
| Amount of explosives per blast [kg] | 345 | 384 | 410 | 534 | 560 | 565 | 600 | 642 | 873 | 914 | 993 | 1011 | 1392 | 1393 | 1462 | 1465 | 1470 |

To determine the seismic effect of group winning blasting, records from two seismic stations were examined. Station A was located approximately 800 m north and station B – 700 m south of the analysed panel. It allows to gather the data from the seismic instruments and observe the vibrations triggered by blasting activities both in the forefield of the mining front (A) and (B) from the side of goafs (mined-out area).

The assessment of the seismic effect included the following elements:
- seismic waveforms analysis in the time domain,
- vibration analysis in the frequency domain using Fast Fourier Transform (FFT),
- Joint Time-Frequency Analysis.

Schemes of the types of signal analysis are shown in Figure 3.

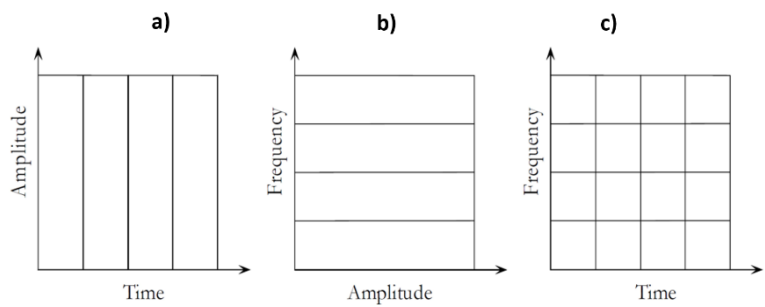

Figure 3. Schemes of analysis in the time (a), frequency (b) and the time-frequency domains (c) [8]

### 3.1. TIME-DOMAIN ANALYSIS

The time-domain analysis is the basic method of assessing seismic vibration intensity that provides information on the type and location of seismic source or energy of event. Moreover, based on the time-domain analysis such information as Peak Particle Velocity (PPV), wavelength and duration of seismic wave can be obtained. In turn, the PPV value can be used for description of the paraseismic vibrations intensity. The maximum value of vibration amplitude can be used for determination of the seismic waves propagation equation. This is therefore the basis for defining i.e. the appropriate amount of explosives in relation to the environmental impact (Winzer et al. 2016). In order to amplify the seismic effect, the expected result of increasing the amount of explosives is to increase the amplitude of rock mass vibrations. All the considered records in the time domain are shown in Figure 4.

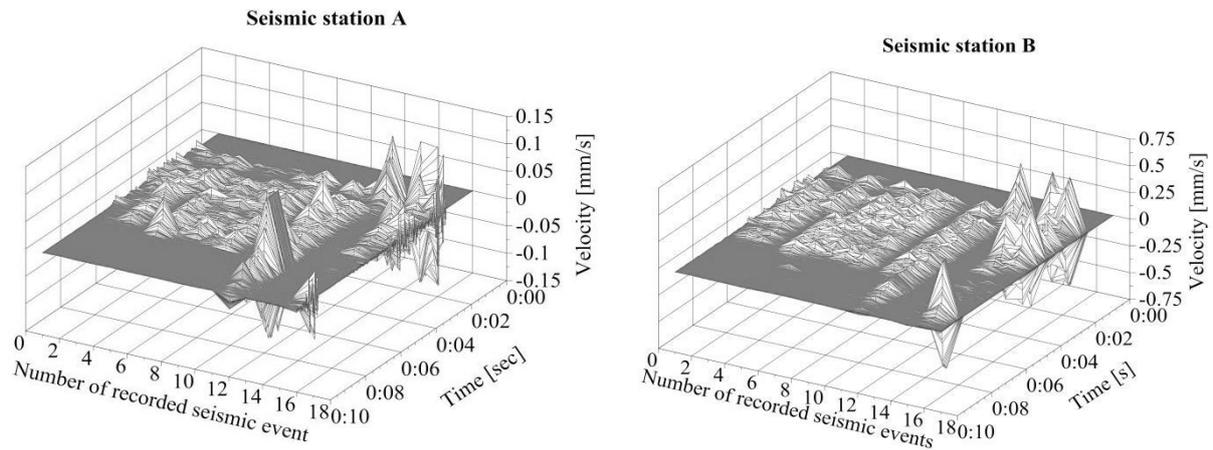

Figure 4. Surface charts of the analysed seismic records

Despite the relatively high variability of the total amount of the explosives fired within each group, only four cases are characterised by significantly higher PPV value. The amplitude distribution over time reveals clear amplitude peaks which are most likely related to the interference of seismic waves generated by the detonation of subsequent delays. In the case of similar amplitudes (cases from 1 to 12), it can be assumed, that induced waves did not interfere, even though the amount of explosives increased. In result, the effectiveness of the provocation wasn't as efficient as it had been assumed. Figure 5 shows the PPV values obtained from both seismic stations.

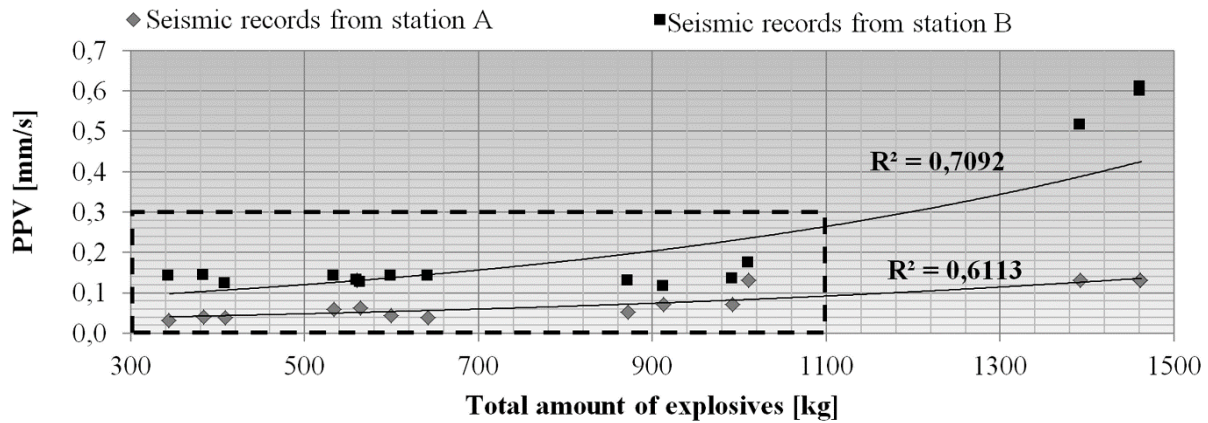

Figure 5. Comparison of recorded PPV values in relation to the amount of the explosives

Based on figure 5 one may conclude, that in the range of applied explosives, i.e. from 345 to 1011 kg (see indicated area) the recorded PPV values remained almost on the same level for both seismic stations, even though the amount of explosives increased three times. It means that the maximum amplitude was not affected by total amount of fired explosives. A significant difference was observed in those blasting operations, in which the total amount of explosives exceeded 1,300 kg. The PPV value recorded at seismic station A, located ahead of the mining panel increased four times and two times at station B, located from the side of goafs. Therefore, it can be assumed that the probability of wave interference triggered by blasting works is increased with a higher total amount of explosives and number of detonated mining faces. However, bearing in mind the accuracy of the delays of detonators used, there is no possibility to obtain constructive/intended wave interference. In this case, the seismic wave amplification had rather a random nature.

### 3.2. FREQUENCY ANALYSIS USING FAST FOURIER TRANSFORM

The analysis of the signal in the time domain only, does not give comprehensive information about the seismic wave characteristics. Each of seismic and paraseismic signals have a complex structure, consisting of many basic harmonic waves, which can be described, for example, by frequencies (Sołtys, 2010). The conversion of waveforms into the frequency domain is mostly performed by mathematical operations, such as a Fast Fourier Transform. Examples of the time history and frequency spectrum, including determination of the dominant frequency, are shown in Figure 6.

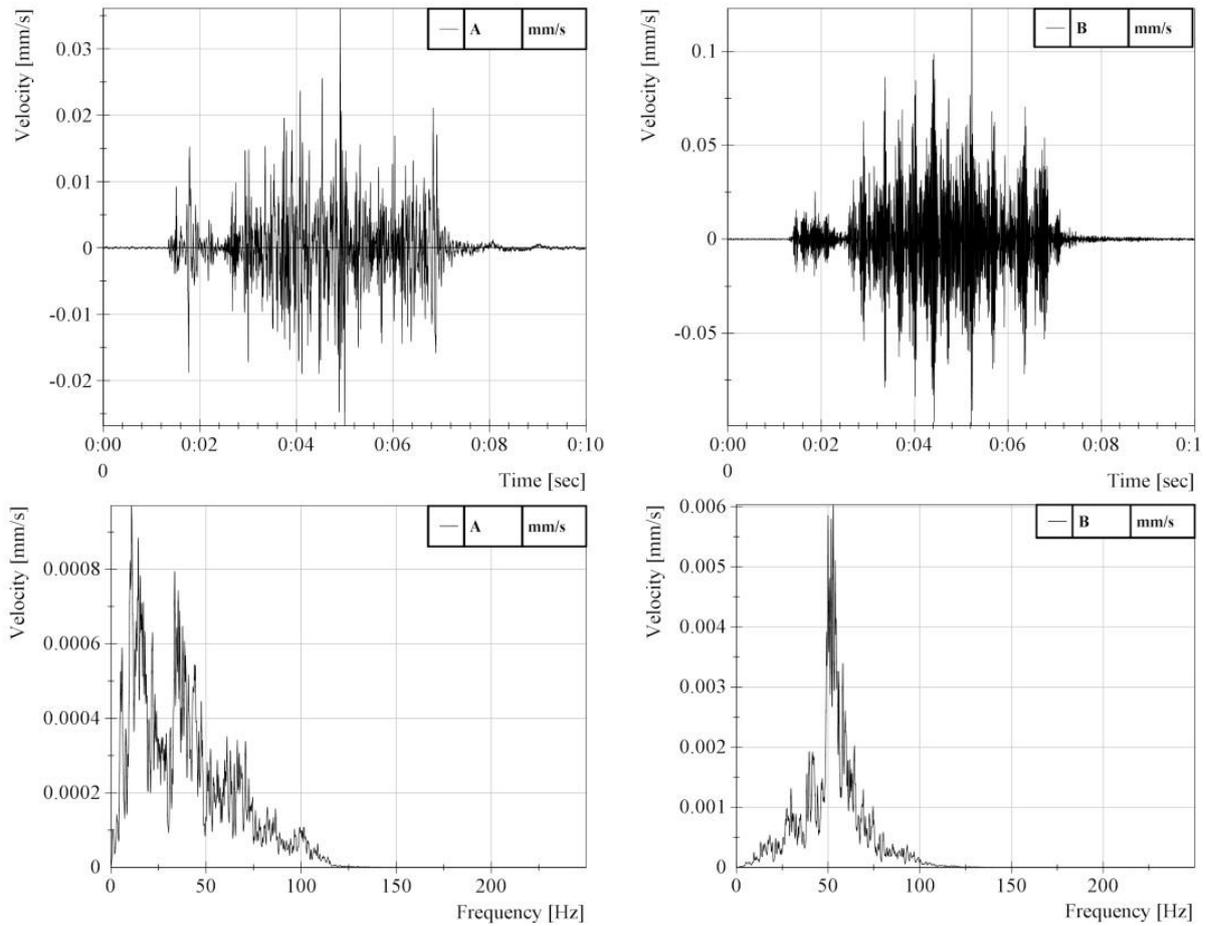

Figure 6. Examples of the time history (top) and frequency spectrum (bottom)

The dominant frequencies were determined using FFT for each particular seismic wave record. Results of calculations, in relation with the total amount of explosives are shown in Figure 7.

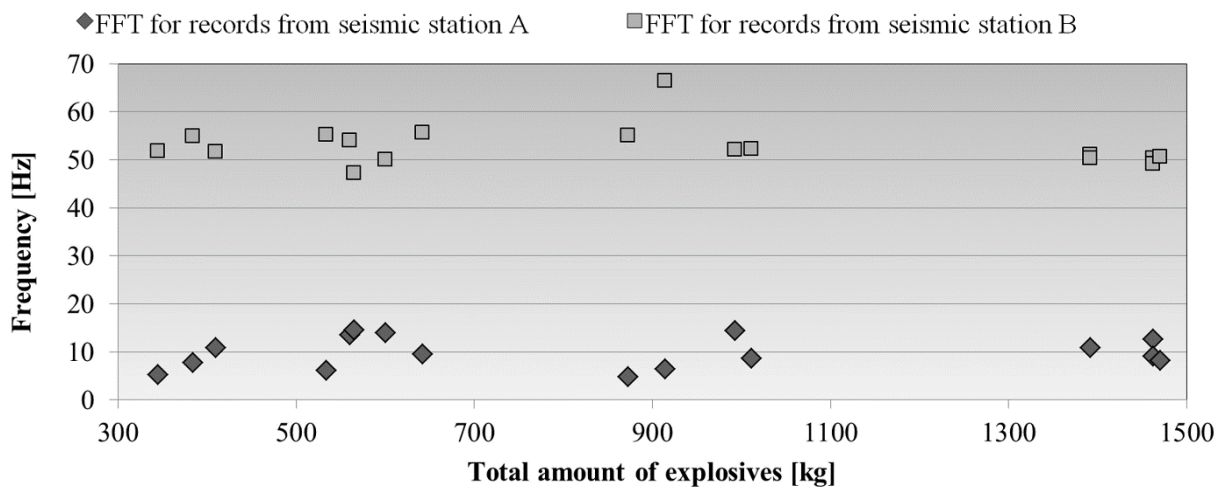

Figure 7. Dominant frequencies calculated for the considered blasting works

As shown in Figure 7, the total amount of explosives as well as the number of simultaneously fired mining faces did not significantly affect the dominant frequency distribution of seismic waves. For the records obtained with the A seismic station, the average value of the dominant frequency was 9.81 Hz with a standard deviation of 3.31 Hz. For the records from the B seismic station, the mean value of the dominant frequency was 52.82 Hz with a deviation of 4.20 Hz. Therefore, it can be concluded, that induced seismic wave frequencies were not related with the amount of explosives.

It should be also noted, that the distance between the mining front line and both seismic stations was similar. However, greater seismic effect was obtained in the forefield of the mining front, where higher amplitudes and frequencies were observed. Probably the reason was higher level of rock mass disintegration from the mined-out area, which damped high frequencies much more than a stressed rocks located in forefield of the mining front.

### 3.3. SHORT-TIME FOURIER TRANSFORM

The time-frequency analysis were carried out using the Short-Time Fourier Transform (STFT). It allows to present the distribution of individual frequencies in time. This method is based on cutting out of subsequent segments of the signal using the window function and then by calculating their Fourier Transform (Sołtys, 2010). It can be applied for the assessment of the effect of multi-face blasting operations to graphically present the energy distribution of the seismic signals in relation to the time and frequency of rock particles vibrations.

Based on calculated spectrograms, it was found, that an increase in the amount of the explosives translates into intensification of the induced seismic waves energy. Furthermore, many cases showed, that simultaneous firing of a higher number of mining faces increases the range of observed signal frequencies. This in turn translates into a greater probability of "matching" to the natural frequencies of the rock mass. In addition, increasing the amount of explosives during group winning blasting affects the high amplitude peaks occurrence.

Selected time-frequency-amplitude charts are shown in Table 2. The axes of the duration, frequency and amplitude of the seismic signals have the same scale range. It allows to highlight the differences in the distribution of the seismic energy generated by considered blasting operations.

Table 2. Comparison of the Time-Frequency-Amplitude plots of selected signals

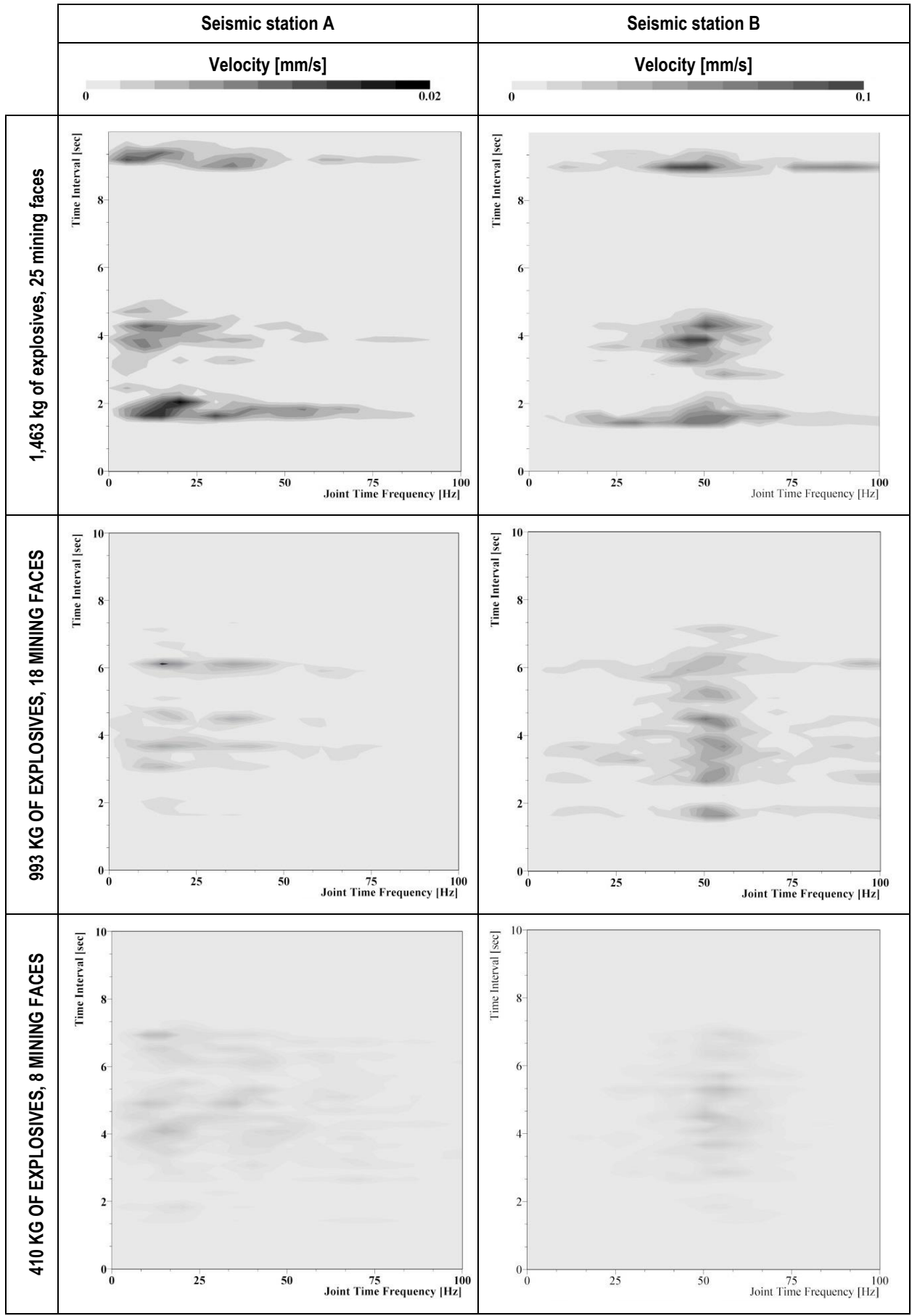

## 4. SUMMARY AND CONCLUSIONS

The purpose of presented study was to determine the seismic effect of multi-face group blasting based on selected mining panel of Polish copper mine in respect to applied rockburst prevention methods. Nowadays, these operations are conducted on a trial-and-errors basis, rather than upon an intentional approach, so they are not effective enough as expected. The influence of the blasting operation on seismic effect was assessed on the basis of detailed analysis of the seismic signals in time, frequency and joint time-frequency domains.

As a result of the analysis one may conclude, that an increase in the amount of applied explosives does not always translate into increase in the PPV value. However, as the number of fired faces increases, the number of high-amplitude peaks increases as well, what is strongly connected with the growth of seismic energy, except the blasting operations in which more than 1,300 kg of explosive were applied. Records of this type of events showed a significant increase of the PPV value, in comparison to the multi-face blasting in which the total amount of explosives varied from 345 to 1,011 kg. The amplitude amplification of the seismic waves was probably associated with seismic waves overlapping from individual mining faces.

Similar conclusions can be drawn from the frequency and time-frequency analyses. Changes of the amount of simultaneous fired explosives does not correlate with the changes in the observed dominant frequencies of the signals. However, when increasing the amount of explosives, the entire frequency range of each particular signal increases.

Based on the analysis of the seismic effect of group winning blasting, it was found, that improvement of rockburst prevention can be achieved by optimising appropriate parameters of blasting operation, i.e. time delays, amount of explosives and number of mining faces.

## ACKNOWLEDGEMENTS

**This paper has been prepared through the Horizon 2020 EU funded project on "Sustainable Intelligent Mining Systems (SIMS)", Grant Agreement No. 730302.**